\documentclass{mn2e}
\title[Infra-red spectra of Seyferts]
{Near-infrared spectra of Seyfert galaxies and line production mechanisms\\}
\author[N. Jackson and R. J. Beswick]
{N. Jackson and R. J. Beswick\\
The University of Manchester, Jodrell~Bank~Observatory, Macclesfield, Cheshire, SK11 9DL U.K.\\}

\newcommand{\fetwo}{[Fe{\sc ii}]}
\newcommand{\othree}{[O{\sc iii}]}
\newcommand{\oone}{[O{\sc i}]}
\newcommand{\ptwo}{[P{\sc ii}]}
\newcommand{\heone}{He{\sc i}}
\newcommand{\paa}{Pa$\alpha$}
\newcommand{\pab}{Pa$\beta$}
\newcommand{\pag}{Pa$\gamma$}
\newcommand{\hb}{H$\beta$}
\newcommand{\ha}{H$\alpha$}
\begin{document}
\input{psfig}
\label{firstpage}
\maketitle

\begin{abstract}
New observations are reported of $J$-band spectra (1.04~$\mu$m --
1.4~$\mu$m) of three Seyfert 2
galaxies, Mkn~34, Mkn~78 and NGC~5929. In each case the spectral range
includes the near-infrared lines of \fetwo, \ptwo, \heone\ and \pab.
Each Seyfert galaxy has a known radio jet, and we investigate the
infrared line ratios of the nuclear and extended regions of each galaxy
compared to the radio structure. In Mkn~34 there is a clear indication
of an extranuclear region, probably coincident with a shock induced by
the radio jet, in which \fetwo\ is considerably enhanced, although the
nuclear emission is almost certainly the result of photoionization by
the continuum of the active nucleus. Similar
effects in extranuclear regions are seen in the other objects, in the
case of Mkn~78 confirming recent studies by Ramos Almeida et al.
A possible detection of extranuclear \ptwo\ emission suggests, if real,
that photoionization by the active nucleus is the dominant line
excitation mechanism over the whole source, including the regions
coincident with the radio jet.
\end{abstract}


\section{Introduction}

Optical and near-infrared studies of nearby active galaxies can give us
a lot of information about the physical processes taking place in these
objects. Seyfert galaxies, as relatively nearby low-luminosity active
galaxies, can be studied in particular detail. These objects contain
emission lines within their spectrum, including in general broad lines
produced within the inner parsec, close to the central black hole and
accretion disk, and narrow lines whose region of production may extend
over a kiloparsec. In addition, many Seyfert galaxies have weak radio
emission, often in a linear structure corresponding to an outflow from
the central active nucleus.

The physics of emission lines is of particular interest. These are
thought mostly to arise through photoionization, and many studies have
been done which use sophisticated photoionization models to reproduce
emission lines in great detail. However, additional physics may be
needed to reproduce some lines. In particular, the near-infrared \fetwo\
line at 1.257$\mu$m has been suggested as a diagnostic of shock excitation,
which may arise as dust grains are dissociated in fast shocks (e.g.
Forbes \& Ward 1993). The idea that shocks may have general
applicability to AGN spectra was argued by Sutherland et al. (2003) and
Dopita \& Sutherland (1995) who succeeded in reproducing many
characteristic photoionized spectra using shock models. This arises
because shocks generate UV and X-ray photons which can then ionize the
gas -- indeed, the difference between the observed properties of such
``autoionizing'' shocks and regions photoionized by a central active
nucleus can be quite subtle and depend on the details of the spectra of
the incident photons.

More recently, Simpson et al. (1996) showed that
the observed \fetwo\ line is also consistent with predominantly being
produced by photoionization, and Mouri, Kawara \& Taniguchi (2000)
conducted more sophisticated calculations suggesting that electron
collisions in a partially-ionized zone associated with a shock may be
the production mechanism rather than grain dissociation. Based on such
studies, Alonso-Herrero et al. (1997) propose the use of the ratio of
infrared \fetwo\ lines to the hydrogen recombination lines to separate
active galaxies from starbursts. Oliva et al. (2001) studied the problem
of shock ionization versus photoionization. They suggested 
that the \ptwo\ line is potentially useful in
disentangling this problem, because similar physical conditions are
needed to produce it to those which produce \fetwo , 
but unlike iron, phosphorus is not produced in
destruction of dust grains. A high Fe/P ratio is thus a prediction of
models in which dust grains are dissociated by shocks. Oliva et al. find
that pure photoionization better fits their spectrum of the very bright
Seyfert galaxy NGC 1068. In fainter galaxies where \ptwo\ is not
detected, even lower limits on the Fe/P ratio may be useful in ruling
out shock models if the spectra are sensitive enough.

Although Seyfert galaxies have been studied extensively in the optical,
high signal-to-noise observations in the near-infrared are rarer. A
compilation has recently been published by Riffel,
Rodr\'{\i}guez-Ardila \& Pastoriza (2006) including many of their own
$JHK$ spectra (Rodr\'{\i}guez-Ardila, Riffel \& Pastoriza 2005). Many of
these spectra suggest that the line ratio of active galaxies have a
\fetwo/\hb\ ratio suggestive of X-ray heating or regions of star
formation. Recently, Ramos Almeida et al. (2006) have published infrared
spectra of one of the objects discussed here, Mkn~78, and deduce from
observations and simulations that the predominant line-excitation 
mechanism in the nuclear region is photoionization by the hard UV
continuum from the active nucleus, although radio-induced shocks may
contribute within one of the radio lobes.

In this work we choose objects in which linear radio structure is
clearly seen in order to investigate the emission-line structure along
the jet axis and look for evidence for shock excitation. If shocks are 
indeed an important part of the physics, the
area where the radio jet is depositing energy into the interstellar
medium is an obvious place to look. The spectra presented here are from
deep exposures (1--2 hours with an 8-m class telescope) and therefore
allow us to investigate both nuclear and off-nuclear emission.

\section{Observations and analysis}

\subsection{Infrared and optical data}

All infrared observations were conducted using the 8-m Gemini-North Telescope, Mauna
Kea, on the nights of 2005 February 17, February 18 and May 22 using the
NIRI imaging spectrograph in f/6 spectroscopy mode. The J grism was
used with a J-band order-sorting filter, 
giving a wavelength range of approximately 1.05$\mu$m --
1.42$\mu$m, together with a 4-pixel slit which corresponds to 
0\farcs46 on the sky and a resolving power of about 600 (Hodapp et al.
2003). Mkn\,78 was observed for 2 hours on 2005 February 17, Mkn\,34 for 
2 hours on 2005 February 18 and NGC~5929 for 70 minutes on 2005 May 22. 
Each observation was divided into 3$\times$30 second
exposures, and each successive exposure was chopped up and down the slit
to make sky subtraction easier and minimize effects of bad pixels. Lamp
flatfield exposures and argon arc exposures were also obtained for Mkn\,34
and Mkn\,78. In the case of NGC\,5929, no argon lamp spectrum is available
and in this case the atmospheric OH lines have been used for wavelength
calibration. 

Data
reduction was performed using the Starlink {\sc figaro} package and
began with combination of the flatfields and division of the data frames
by a flatfield normalised in the spectral direction. 
A second-order
polynomial fit to at least six arc lines was obtained which gave a
maximum error of 0.1~nm across the wavelength range. 
The spectra were
corrected for S-distortion using three spectra of the Hipparcos stars
taken at different positions on the chip. The sky was then
subtracted using the Figaro task {\sc polysky} and the spectra extracted
using the optimal extraction algorithm of Horne (1986) using the Figaro
task {\sc optextract}.

Relative flux calibration was performed assuming a blackbody spectrum of
7200K for the Hipparcos F0V star HIP36366 (Perryman 1997) and temperatures
of 7200K and 7000K for the stars HIP56601 and HIP75411 respectively. The
Pa5$\rightarrow$3, 6$\rightarrow$3 and higher order absorption bands were
interactively edited out of the stellar spectrum before this process in
order to avoid them appearing as emission lines in the divided spectrum.

For deriving velocity profiles, successive rows of the sky-subtracted
image have been extracted, continuum-subtracted using line-free regions
immediately around each line, and fitted with Gaussian profiles. Profile
fitting  has been done by software written by us, using the Levenberg-Marquardt
algorithm as implemented by Press et al. (1992). 

In addition to the infra-red data, some data from the Isaac Newton
Group (ING) archive was extracted for the Seyfert galaxy Markarian~34.
These data were taken using the William Herschel 4.2-m telescope on La
Palma on 1998 December 17 using the ISIS two-arm spectrograph. On the 
red arm of this spectrograph the TEK-2 chip was used together with an R600R
grating, using a 1\farcs23 slit and a resulting spectral resolution of just 
over 0.2nm. The data
were co-added and fluxes of \oone\ and \ha\ derived assuming a flat
spectral response over this wavelength range, because both the TEK-2 chip
and the R600R grating have a flat spectral response over this range to 
within a few percent. 

\subsection{Radio data}

Complementary, mainly previously-published (Falcke, Wilson \& Simpson 1998; Whittle \& Wilson 2004)
public domain radio data for each of these three sources were obtained and re-mapped
for comparison purposes from the NRAO's Very Large Array (VLA) archive. In each
case these data were observed at a wavelength 3.6\,cm and in the VLA's
highest resolution configuration, providing an angular
resolution of $\sim$0.2-0.3 arcsec which is directly comparable to the spectroscopic
observations presented. A brief summary of the observing
parameters is given in Table\,\ref{tab1}. Each data-set was calibrated
using standard data reduction techniques within NRAO's {\sc aips}
package where these data were edited before phase solutions from a nearby
reference source were applied. In each case the flux density scale was
calibrated with respect to 3C\,286 using the Baars {\it et al.} (1977)
scale.

\begin{table*}
\centering
\caption{Summary of archival VLA observations.}
\label{tab1}       
\begin{tabular}{lcccl}
\hline\noalign{\smallskip}
Source&Date&Phase ref. & Time on source & Original publication \\
&&&&\\
\hline
Mkn\,34&1996 Nov 4&1030+611&166 min & Falcke, Wilson \& Simpson 1998\\
NGC\,5929&2003 Jun 20&1506+426&8 min&\\
Mkn\,78&1990 Apr 16&0716+714&340 min&Whittle \& Wilson 2004\\

\noalign{\smallskip}\hline
\end{tabular}
\end{table*}

\begin{figure}
\psfig{figure=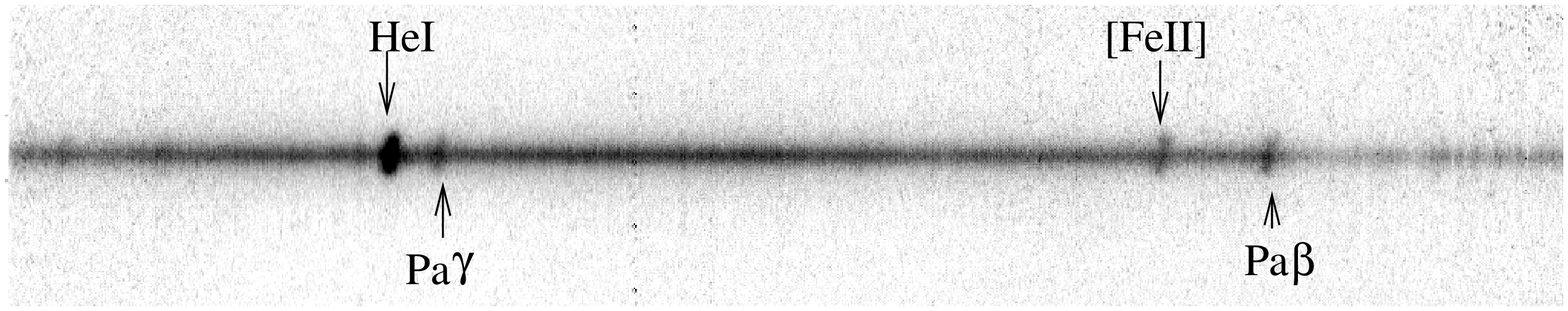,width=9cm}
\psfig{figure=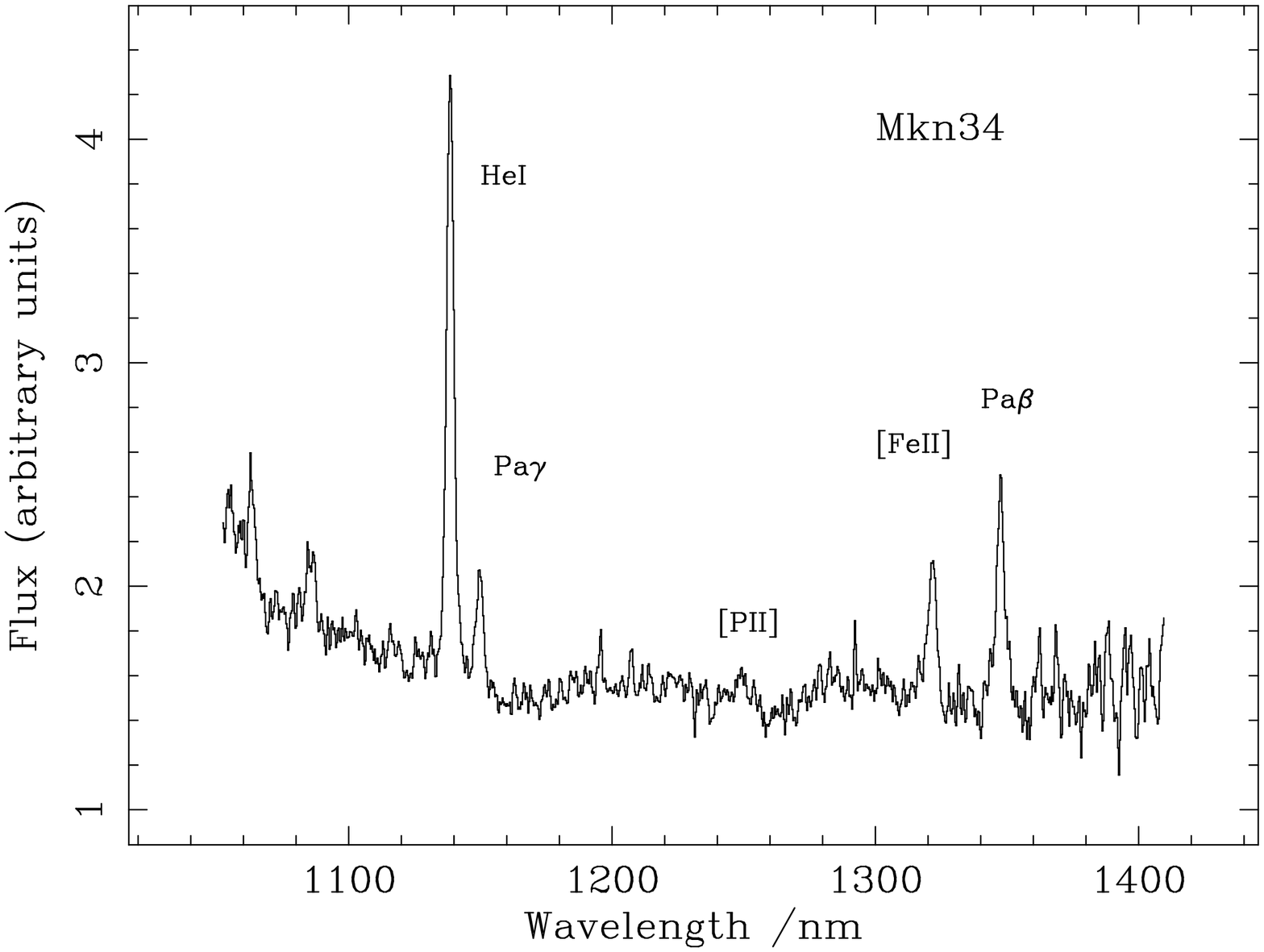,width=9cm,angle=-90}
\caption{$J$-band spectrum of Mkn~34, showing the brightest emission
lines.}
\end{figure}

\section{Results and line diagnostics}

\subsection{Mkn 34 ($z$=0.0505)}

Markarian 34 is a Seyfert 2 galaxy with a 2\farcs4 extended radio
structure whose extended optical line emission has been studied in
detail by Whittle et al. (1988) and Falcke, Wilson \& Simpson (1998). 
Its off-nuclear line emission shows a
velocity profile with relatively broad \othree\ emission extending
slightly beyond the radio lobes. There is a distinct difference between the
velocity of the line emission near to the nucleus and that beyond the
radio lobes. The new infra-red spectrum was 
taken with the slit at
a position angle 158$^{\circ}$ East of North, coincident with the radio
axis. Figure 1 shows the image of the spectrum along the slit, showing a
clear rotation-type velocity profile in all securely detected lines (\heone, \fetwo,
\pab\ and \pag).

\ptwo\ is detected at a marginal level in the nuclear spectrum. In the
nuclear region, the ratio \fetwo$\lambda$1.257/\ptwo\ is 3.2$\pm$1.0,
which is consistent with the range of values typical of nuclear
photoionization (Oliva et al. 2001).


The radio map of Falcke et al. (1998) shows structure similar to that seen
earlier by Ulvestad \& Wilson (1984) at 6 and 20\,cm, but with
higher angular resolution and sensitivity shows the
definition of the radio jet structure more clearly. In particular the
jet is resolved into several discrete knots and a significant change
in the projected position angle of the tip of the southeastern jet is
seen.  

\begin{figure}
\psfig{figure=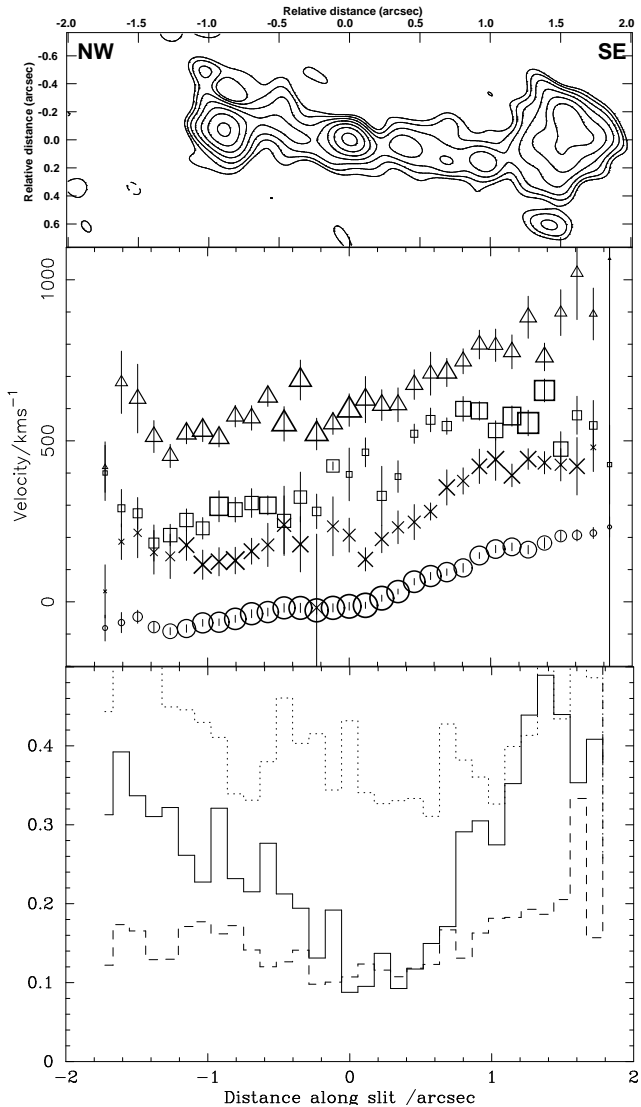,width=8.5cm}
\caption{Central panel: Velocity profile of Mkn\,34. Four lines are shown, \heone\
(circles), \pag\ (crosses), \fetwo\ (squares) and \pab\ (triangles).
Line quantities have been measured by Gaussian fits to successive rows
of the spectral image, and the amplitude within each sequence of line
fits is proportional to the size of the symbol. Each successive line has
been offset by 200~km\,s$^{-1}$ for clarity. Note the clearly different
pattern of the \fetwo\ emission from the other lines; it is highly
concentrated in region of the SE radio knot. The radio map (top panel)
from data presented by Falcke et al. (1998) is shown to the same scale, 
and rotated to the axis of the spectrograph slit. The bottom panel shows
the line ratios compared to \heone\ of \fetwo\ (solid line), \pag\
(dashed line) and \pab\ (dotted line).}
\end{figure}

The most striking feature of the spectrum is the different distribution
between \fetwo\ and the other lines. Figure 2 illustrates the rotation
curve and strength of the lines, and it is clear that \fetwo\ has a
bright knot about 1\farcs2 to the SE. This is just behind 
the bright radio knot in the southeastern jet imaged here and by Whittle et al.
(1988). Although the Paschen recombination lines are slightly enhanced
in the off-nuclear regions compared to the nucleus, the ratio of 
\fetwo/\pab\ becomes a factor of 2--3 stronger away from the nucleus,
suggesting that different ionization mechanisms may be operating here.
The \heone\ line displays a different behaviour to the hydrogen
recombination lines, showing a smooth decrease in each direction away
from the nucleus with little sign of any increase in strength in the
radio lobes.

\begin{figure}
\psfig{figure=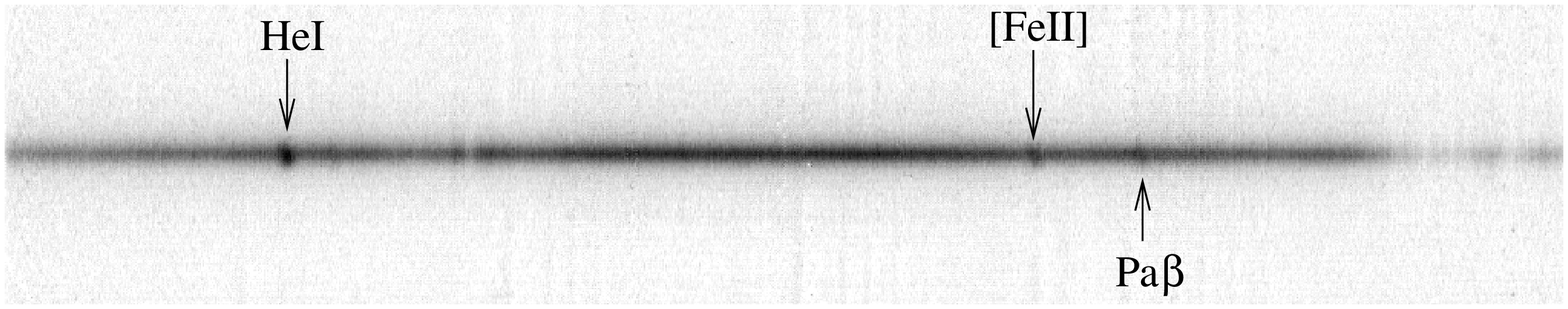,width=9cm}
\psfig{figure=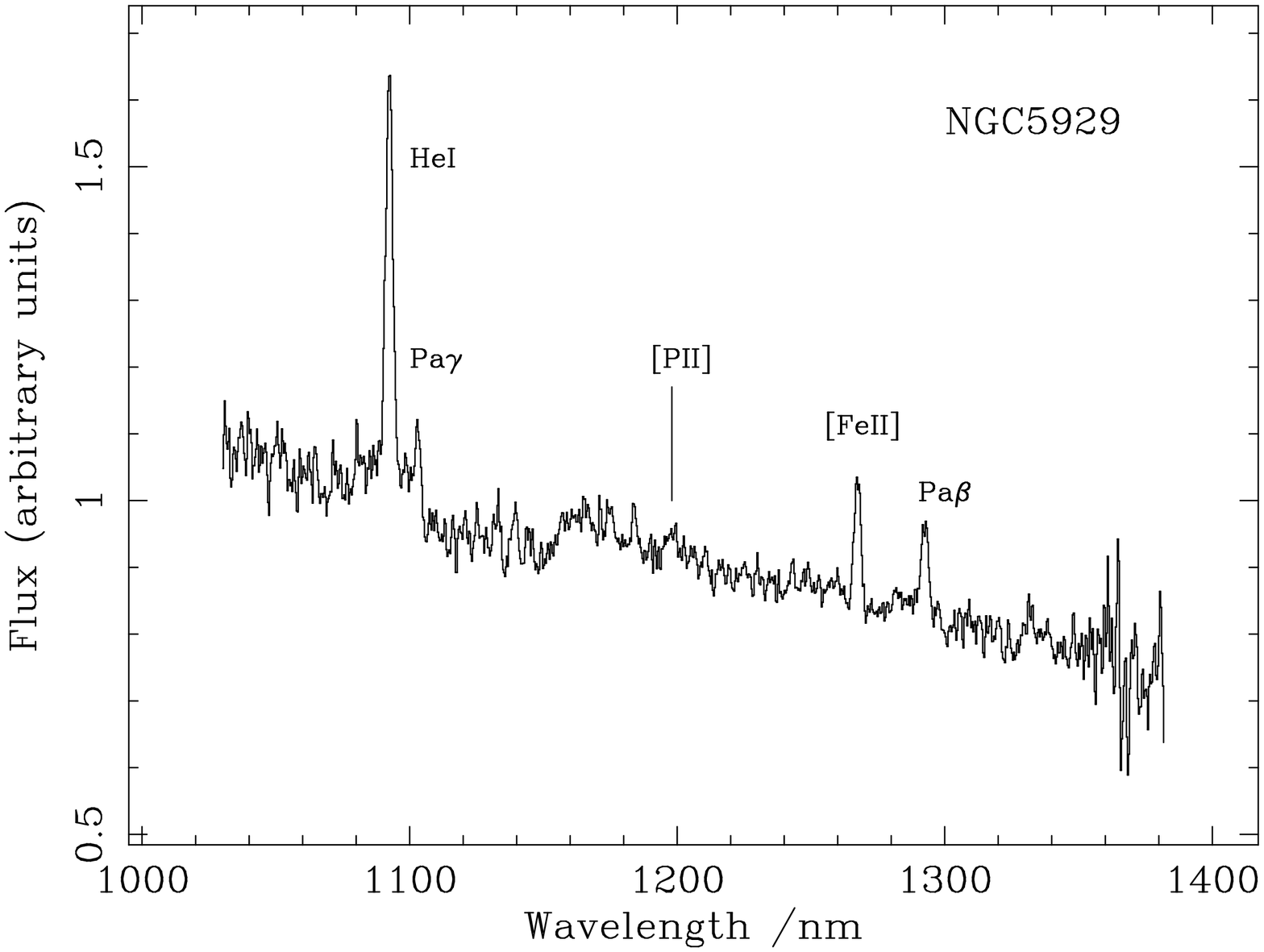,width=9cm,angle=-90}
\caption{$J$-band spectrum of NGC~5929, showing the brightest emission lines.}
\end{figure}

\subsection{NGC\,5929 ($z$=0.00831)}

NGC5929 is a Seyfert galaxy with a radio jet extending about 0\farcs7
in PA $\sim 60^{\circ}$ either side of a compact nucleus (Su et
al. 1996; Cole et al. 1998)
and extended optical emission partly coincident with the western radio lobe
(Bower et al. 1994),
strongly suggesting an interaction. Nuclear infrared spectroscopy was
performed by Simpson et al. (1996) who clearly detected the \fetwo\ line
and found it to be marginally resolved. 
The current observations have a smaller pixel scale, and we clearly see the resolution 
in this object, using a slit at the position angle of the
extended optical emission previously seen (65$^{\circ}$). The nuclear
spectrum is shown in Figure 3.

\begin{figure}
\psfig{figure=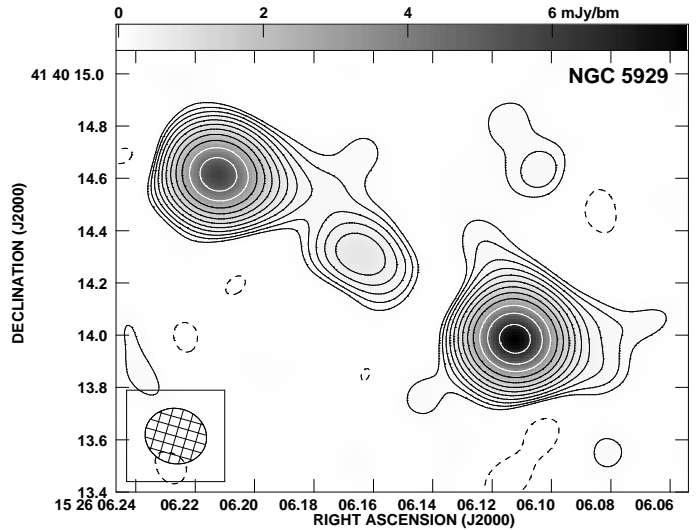,width=9cm,angle=-90}
\caption{VLA 3.6cm image of NGC\,5929, showing features similar
to the map of Su et al. (1996). The map is contoured at
multiples of $\sqrt 2$ $\times$ 0.146\,mJy\,bm$^{-1}$
($3\sigma$). The image has been convolved with Gaussian beam of
0\farcs234$\times$0\farcs211 with a of PA=74$\degr$.}
\end{figure}

In the nuclear emission the \ptwo\ line is detected at just over the
3$\sigma$ level; a previous spectrum reported by Rodr\'{\i}gues-Ardila,
Riffel \& Pastoriza (2005) and Riffel et al. (2006)
did not detect this line. A Gaussian fit to the \fetwo\ and \ptwo\ lines was
carried out, constraining the \ptwo\ line width to be equal to that of
\fetwo . The \fetwo/\ptwo\ ratio is 4.3$\pm$1.1 in the nuclear region,
which again is typical of photoionized regions and considerably less
than the value of $\sim$20 expected for shock ionization (Oliva et al.
2001). The \fetwo/\pab\ ratio is approximately 1, in agreement with the
previous spectra.

\begin{figure}
\psfig{figure=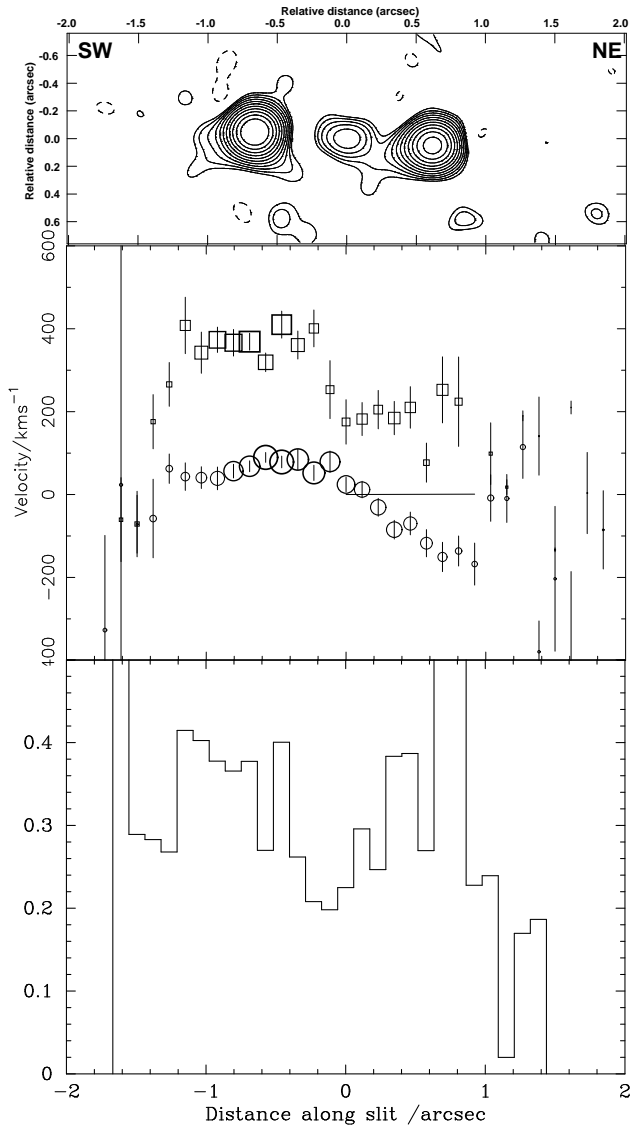,width=8.5cm}
\caption{Central panel: Velocity profile of NGC~5929. Two lines are shown, \heone\
(circles) and \fetwo\ (squares).
Line quantities have been measured by Gaussian fits to successive rows
of the spectral image, and the amplitude within each sequence of line
fits is proportional to the size of the symbol. Each successive line has
been offset by 200~km\,s$^{-1}$ for clarity. The radio map (top panel)
is the same as Fig.\,4,  but is shown to the same scale, and rotated to the axis of the
spectrograph slit. The bottom panel shows the variation of the line
ratio \fetwo /\heone over the same scale.}
\end{figure}

Figure 4 shows the radio map; similar structure is seen to the
earlier radio map of Su et al. (1996) and Cole et al. (1998) with a
compact core and two lobe structures detected. 

The most remarkable feature of the spectrum is the wholesale displacement of the
line emission from the centre of the galaxy and of the velocity profile
(Figure 5).
This effect is extreme around the southwestern radio lobe, where the 
\fetwo\ line strength peaks. The \heone\ line peak is also offset, 
although by less than the \fetwo . Optical studies with the {\it Hubble Space
Telescope} (Bower et al. 1994) have
also revealed \othree\ emission which peaks around the southwestern
radio lobe. Around the northeastern radio lobe,
there is some evidence of disturbed kinematics in the \fetwo\ line.

\subsection{Mkn\,78 ($z$=0.03715)}

Mkn 78 has recently been the subject of an extensive optical study with
the Hubble Space Telescope by
Whittle \& Wilson (2004), Whittle et al. (2004) and Whittle et al. (2005)
specifically directed at disentangling shock ionization and
photoionization. This Seyfert galaxy has extended radio emission, in a
linear structure about 2$^{\prime\prime}$ long in an E-W direction, and Whittle
\& Wilson (2004) have imaged a large and complex extended structure in
\othree\ extending 3$^{\prime\prime}$ from the nucleus in each direction. In
these observations the slit was aligned in PA 90$^{\circ}$ in the hope
of detecting significant extended structure in the infrared emission
lines. The nuclear spectrum of this object is shown in Figure 6, and the
measurements of the extended emission and velocity profile are displayed
in Figure 7.

\begin{figure}
\psfig{figure=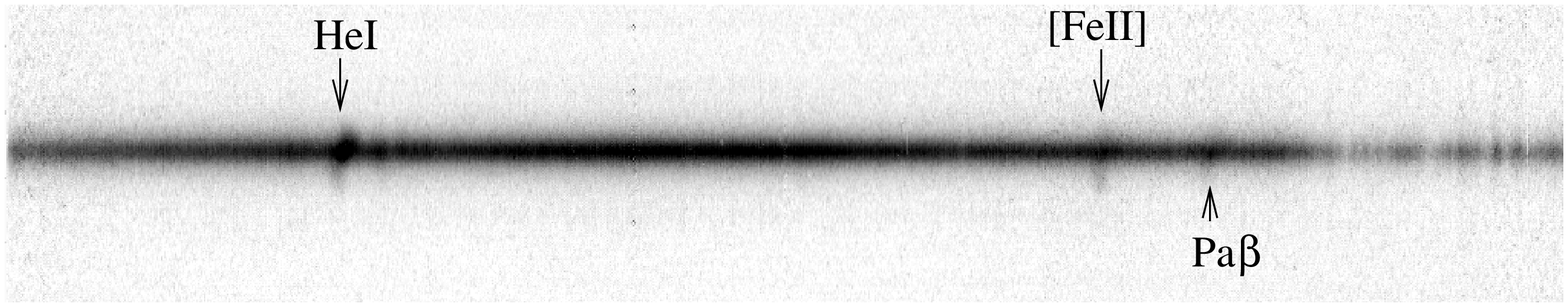,width=9cm}
\psfig{figure=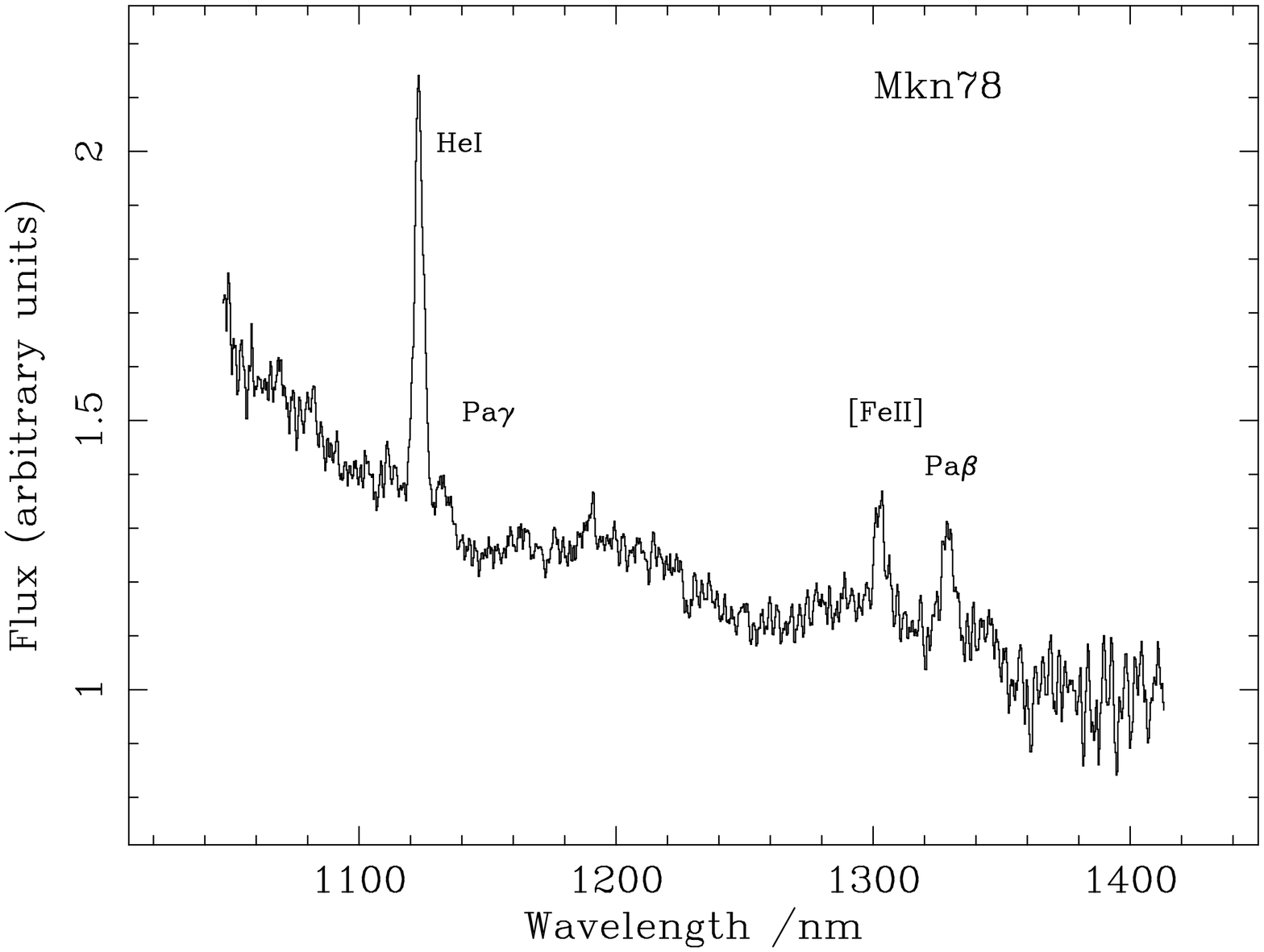,width=9cm,angle=-90}
\caption{$J$-band spectrum of Mkn~78, showing the brightest emission lines.}
\end{figure}


In the optical observations of Whittle et al. (2005), very complex 
extended optical line emission was detected in all major recombination
and forbidden lines. In these infrared observations, the structure is not well
resolved either spatially or spectrally. However, the overall rotation
curve is detected in extended emission in three major emission
lines, \heone, \fetwo\ and \pab, covering about 2$^{\prime\prime}$ to the East
of the nucleus. Relatively little line emission is seen to the West,
which is also the direction of lower surface brightness in \othree\ from
the maps of Whittle \& Wilson (2004). 

A recent detailed infrared study has been carried out with the LIRIS
spectrograph on the William Herschel Telescope by Ramos Almeida
et al. (2006). They find a nuclear ratio of \fetwo /\paa\ which is
consistent with photoionization from the active nucleus. Moreover, they
detect \ptwo\ in the nucleus at a strength that is a factor of 1.82 lower than
the \fetwo$\lambda$1.644 line. This ratio is also consistent with
photoionization, being similar to the ratio found in NGC\,1068 by Oliva
et al. (2001), and is inconsistent with shock models. However, in the
extended emission associated with the western radio lobe, they find that
the \fetwo\ line  strength increases by a factor $\sim$2 relative to the
hydrogen recombination lines, suggesting a contribution from interaction
with the radio bow shock. We find a similar effect in our data, with an
enhancement in \fetwo\ which peaks at $\sim 0\farcs6$ from the nucleus.
However, the association of increased \fetwo\ emission with the radio
structure is {\em less} obvious in this object than in the other two in
this study.  In the western lobe the peak of the \fetwo\ does
coincide with the W-knot of radio emission seen at higher resolution
by Whittle et al. (2004) and shown in the superposed uniformly
weighted image in the top panel of figure 7. The stronger, eastern radio jet is
co-spatial with a region in which the \fetwo\ (and \pag) emission
becomes extremely faint and difficult to measure accurately. The
\heone\ line displays different behaviour; as in Mkn\,34, its strength
decreases smoothly away from the nucleus. The \pag\ line, although it
can be followed for about 2$^{\prime\prime}$ either side of the
nucleus, is not detected at high enough signal-to-noise to describe
its structure in detail.


\begin{figure}
\psfig{figure=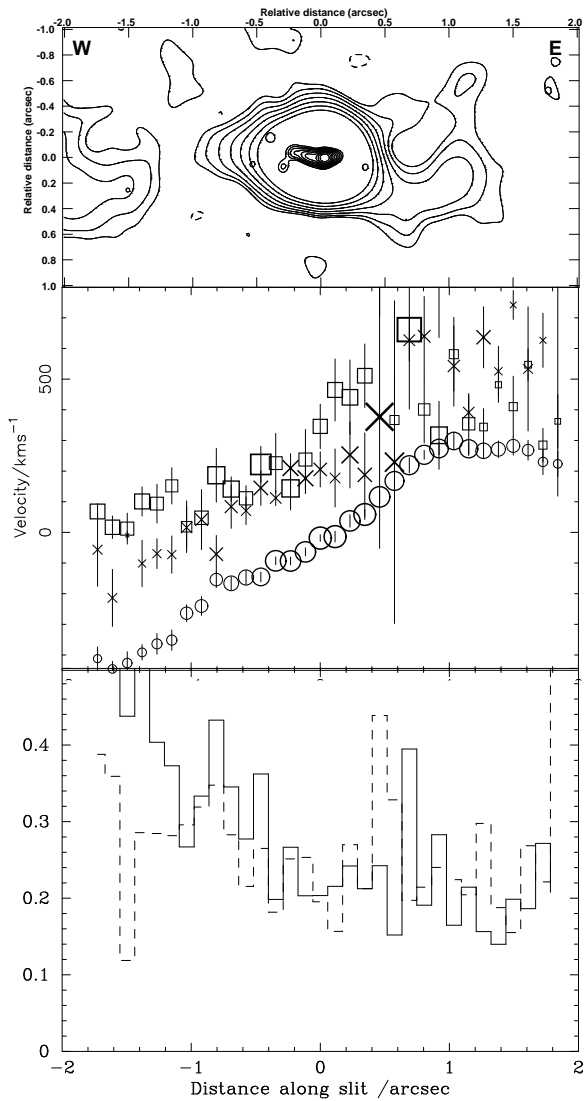,width=8cm}
\caption{Velocity profile of Mkn 78. Three lines are shown, \heone\
(circles), \pab\ (crosses), \fetwo\ (squares).
Line quantities have been measured by Gaussian fits to successive rows
of the spectral image, and the amplitude within each sequence of line
fits is proportional to the size of the symbol. Each successive line has
been offset by 200\,km\,s$^{-1}$ for clarity. The radio map (top panel) is 
made using the data of Whittle \& Wilson (2004); it
shows the central part of the radio emission, rotated to the axis of the
spectrograph slit. Superposed on this radio map is a higher resolution
(0\farcs075) uniformly weighted radio image of the core and inner
jet. This inner image uses the same contouring scheme as the outer
image with a lowest contour value contour of 0.1\,mJy\,beam$^{-1}$.  
The apparent high amplitude of the \fetwo\ line at
0\farcs7 is the result of a noisy fit which is unlikely to be
representative of the true flux; the same is true of the \pab\ line at
0\farcs45.
The bottom panel shows the line ratios compared to \heone\ of \fetwo\ (solid line),
and \pab\ (dashed line).}
\end{figure}

\section{Ionization mechanisms and diagnostics}

\subsection{Nuclear emission: the \fetwo/\ptwo\ ratio}

This paper adds to the total number of detections of
\ptwo$\lambda$1.188$\mu$m in the literature. Since its original
detection in NGC~1068 by Oliva et al. (2001) the line has been detected
in the Seyfert 1 galaxy Akn~564 by Rodr\'{\i}guez-Ardila, Riffel \& Pastoriza
(2005), in
Mkn~78 by Ramos Almeida et al. (2006), and in a total of 11 galaxies in
the compilation of Riffel et al. (2006). Although very marginal, the
detections of \ptwo\ in Mkn~34 and NGC~5929 suggest \fetwo/\ptwo\ ratios
of 3--4, in agreement with ratios determined in other objects.
Comparison with the models presented by Oliva et al. (2001) suggests
that the dominant ionization mechanism is photoionization by the active
nucleus rather than shock excitation; fast shocks would be expected to
disrupt dust grains and produce line ratios a factor of 10 higher.

\subsection{Extended emission}

\subsubsection{Line diagnostic diagrams}

There have been a number of investigations which have attempted to
compare near-infrared 
line ratios with ionization codes. Many have used the {\sc cloudy}
code (Ferland et al. 1998) for a range of physical parameters, namely:
density of hydrogen, ionization parameter, ionizing power-law slope,
metal abundance and presence or absence of dust grains (Simpson et al.
1996; Ramos Almeida et al. 2006). Others have used the {\sc mappings} code or
modifications to it (Sutherland et al. 1993; Mouri et al. 2000). Of
particular interest is the ratio of \fetwo\ to the other lines available
in the near-infrared spectrum, in particular the hydrogen recombination
lines. Also interesting is the optical \oone$\lambda$6300 line, as this has a similar
ionization potential to \fetwo\ and is produced in a partially-ionized zone
with similar physical properties to the region producing \fetwo . It 
can therefore serve as an alternative
diagnostic of shock excitation to \ptwo . Although much stronger than
\ptwo , \oone\ suffers from the disadvantage that the \oone /\fetwo\
ratio is subject to substantial reddening and is difficult to measure
with the same instrument.

The effect of different physical parameters on the relative strengths of
the \fetwo\ and hydrogen recombination lines (in particular \pab\ ) can
be summarized as follows (Simpson et al. 1996, Mouri et al. 2000; Ramos
Almeida et al. 2006):

\begin{itemize}
\item Metal abundance changes the \fetwo/\pab\ ratio by approximately
the scaling of the abundance, for example by a factor of $\sim$6 from
solar to Orion abundance.
\item Neutral hydrogen density has relatively little effect, except at
very high densities when the forbidden lines are in any case close to
collisional de-excitation. For most values of ionization parameter,
increasing densities decreases the relative strength of \fetwo\ very
slightly.
\item Increasing ionization parameter causes a slow decrease in the
\fetwo/\pab\ ratio below $U=10^{-1.5}$, and a rapid increase above this.
\item The slope of the power-law has a drastic effect on the 
\fetwo/\pab\ ratio; a steepening by 0.4 (from $-$1.0 to $-$1.4) 
typically halves the ratio, and the ratio becomes even smaller if a
40000-K blackbody model is adopted instead. This was considered in
detail by Simpson et al. (1996), although Ramos Almeida et al. (2006)
use an index of $-$2.0 throughout their simulations.
\end{itemize}

In order to compare results with physical models and extract physical
information from the line ratios, we have used the {\sc cloudy} program
(Ferland et al. 1998), version c06.02b. Models have been calculated for
ionization parameters ranging from $10^{-3.5}$ to $10^{-1.5}$ in steps
of $10^{0.5}$ and electron
densities between $10^{3.5}$ and 10$^6$, also in steps of $10^{0.5}$. 
For each combination of ionization parameter and density, two
different metallicities ($0.4Z_\odot$ and $Z_\odot$) and two values of
the spectral index of the power-law photoionizing continuum ($-1$ and
$-2$) have been used. Grains were added to the simulation and were
adjusted to be similar to Orion grains.

\begin{figure}
\psfig{figure=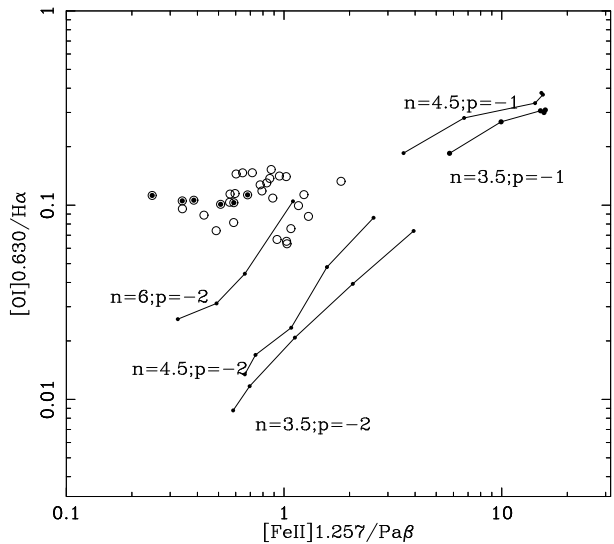,width=8cm,angle=-90}
\caption{Diagnostic diagram of \fetwo$\lambda$1.257/\pab\ vs. \oone/\ha\
compared to the observations of 
Mkn~34. Observational values (circles) within 2$^{\prime\prime}$ either 
side of the nucleus are shown, and pixels $<$0\farcs4 from the nucleus
are shown as solid blobs surrounded by a circle. Model tracks from the
{\sc cloudy} program are also shown, for various values of the ionizing
continuum power-law slope ($p$), hydrogen density ($n$, in log$_{10}$cm$^{-3}$).
All tracks are calculated for $Z=0.4Z_{\odot}$. Within each track, the
ionization parameter $U$ varies from $-$1.5 to $-$3.5, with the higher
values of \fetwo$\lambda$1.257/\pab\ occurring at lower $U$.}
\end{figure}

The diagnostic diagram of \fetwo$\lambda$1.257/\pab\ vs. \oone/\ha\ is
shown for Mkn~34 in Figure 8. This is the same diagram as presented in
Figure 3 of Mouri et al. (2000). The trends in line ratio with ionization
parameter, hydrogen density and slope of the ionizing continuum agree
with previous determinations. In particular, an increase in ionization
parameter causes decreases in both the \fetwo$\lambda$1.257/\pab\ and the
\oone/\ha\ ratio of about the same magnitude, provided that the slope of
the ionizing continuum, is relatively steep ($<-1.5$); in practice,
relatively steep continua are needed to fit these data.

We have already noted that the ratio between \fetwo$\lambda$1.257 and
the hydrogen recombination lines increases in the objects studied as one
moves from the nuclear regions to the areas associated with radio lobe
emission. In principle, if the sole mechanism operating is ionization by
hard-UV radiation from an active nucleus, this change in line ratio 
could be caused by a decrease in ionization parameter resulting from a
geometrical dilution in the ionizing photon flux or an increase in the
electron density associated with compression by the radio jet. This is 
indeed the conclusion of Falcke et al. (1998) who study the variation 
of the \othree/\ha\ ratio and find that it decreases away from the
nucleus. In this case, however, one would also expect the \oone/\ha\ 
ratio to increase, and this does not seem to be the case
(Figure 8). In fact, this ratio is actually
smaller in the southeastern lobe of Mkn~34, where the strongest
interaction may be taking place, than in the nucleus of Mkn~34.
Since the \oone\ and \fetwo\ lines are produced in the same type of
partially ionized zone and in similar physical conditions, the lack of
correlation between their ratios to hydrogen recombination lines is
puzzling. 

The same, however, does not appear to be the case in Mkn~78. Whittle et
al. (2005)'s extensive study of the extended optical emission lines in
this object suggest that all lines, including \oone , follow standard
tracks from photoionization diagnostics, with a variation of about a
factor of 10 in $U$; this is sufficient to produce a difference of about
a factor of 2 in \fetwo$\lambda$1.257/\pab .

Images of some optical emission lines in the extended regions of NGC~5929
are presented by Bower et al. (1994) and discussed further by 
Su et al. (1996); however, we do not have enough
information on this object to present detailed diagnostic diagrams
comparable with previous work.

\subsubsection{Tentative detections of extended \ptwo }

A definitive test would be the detection of the \ptwo\ line in the
extended emission line region. In Figure 9 an extraction is shown of
0\farcs7 of the slit in the region where the \fetwo\ line is strongest,
just behind the southeastern radio lobe of Mkn~34. The \fetwo\ line in
this region has been fitted with a Gaussian, and is shown shifted to the
\ptwo\ wavelength and divided in intensity by factors of 4 and 20. There
is a spike at exactly the expected wavelength, although it is extremely
close to the noise level. If this is real, it clearly indicates that the
mechanism operating in this extended region is photoionization, with
little or no shock contribution. Unfortunately the signal-to-noise in
this spectrum is just too low to make a definite statement.

\begin{figure}
\psfig{figure=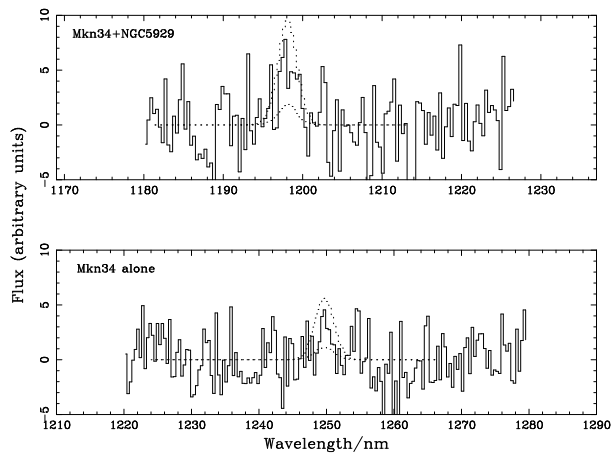,width=8cm,angle=-90}
\caption{(Bottom panel) Area around the \ptwo\ line in the six pixels 
(0\farcs7) near the
southeastern radio lobe of Mkn\,34 with the strongest \fetwo\ emission.
The {\em predicted} \ptwo\ emission, deduced from scaling the \fetwo\
emission by the known wavelength ratio and by scaling in intensity by
factors of 4 and 20, is shown by dotted lines. The wavelength scale is
that of the observed frame in Mkn\,34. (Top 
panel) Combined extranuclear emission from near to the southeastern radio lobe
of Mkn\,34, as above, and the region of 0\farcs6 
around the southwestern radio lobe of NGC\,5929.
Both of these regions have enhanced \fetwo\ emission. The spectra have
been coadded and corrected to the same wavelength scale by use of the
\fetwo\ line. The predicted position of the \ptwo\ line is shown in
each case; the dotted lines show the \ptwo\ line as it would appear if it
were factors of 4 and 20 weaker than \fetwo . The wavelength scale is
that of the observed frame of NGC\,5929.}
\end{figure}

However, repeating this exercise in the extended region of NGC\,5929
associated with the southwestern radio lobe (Figure 5) and in which the
\fetwo\ is also enhanced over its value in the nuclear region, reveals a
similar {\em very} marginal \ptwo\ detection in the extended emission. Adding
the two together, with the appropriate scaling in wavelength (Figure 9)
gives a more secure detection of the extended \ptwo\ at a level of about
0.2 times the strength of the extended \fetwo\ line. This suggests,
somewhat surprisingly, and despite the association with the radio jet, 
that the dominant mechanism of line excitation in
these regions is still photoionization rather than shock excitation.

\section{Conclusions}

A number of near-IR studies of Seyfert galaxies with extended optical
emission have now been carried out, in particular Mkn\,78 (Ramos Almeida
et al. 2006 and this work), Mkn\,34 and NGC\,1068 (Oliva et al. 2001). The
picture which emerges with reasonable consistency is that the nuclear
regions are almost certainly dominated by photoionization by UV photons,
consistent with a mechanism involving the active nucleus. This is
indicated both by the ability of photoionization to reproduce observed
emission line ratios, and specifically by the inconsistency of the
\fetwo/\ptwo\ diagnostic with the enhanced ratios expected from shock
models. In the extended emission regions which coincide with radio
jets, there is a consistent pattern of changes in emission line ratio
consistent with compression of the gas by the bowshock associated with
radio emission. Tentative detections of \ptwo\ in these regions,
however, suggest that photoionization is still the dominant mechanism.
The fact that the peak line emission is slightly behind the head of the
radio jet in Mkn\,34 may suggest that the gas has been compressed by the
shocks but ionized by the active nucleus after the \fetwo\ has been
returned to the gas phase, rather than by UV and X-ray photons from the shocks
themselves. In this case the fact that NGC\,5929's peak of \fetwo\ is
coincident with the radio lobe could indicate that the shock compression
of the gas has occurred more recently. 
It is important to confirm these determinations in other objects, using
longer integrations on large (8-m class) telescopes.

\section*{Acknowledgments}

Based on observations obtained at the Gemini Observatory, which is operated by the
Association of Universities for Research in Astronomy, Inc., under a cooperative agreement
with the NSF on behalf of the Gemini partnership: the National Science Foundation (United
States), the Particle Physics and Astronomy Research Council (United Kingdom), the
National Research Council (Canada), CONICYT (Chile), the Australian Research Council
(Australia), CNPq (Brazil) and CONICET (Argentina).  RJB acknowledges financial support from 
the European Commission's I3 Programme ``RADIONET'' under contract No. 505818.
 The VLA is operated by the National Radio Astronomy
Observatory which is a
facility of the National Science Foundation operated under cooperative
agreement by Associated Universities, Inc. We thank the
observers, I. Song and S. Knights and the operators, B. Walls and G.
Trancho, for carrying out the observations.

\section*{References}

\noindent Alonso-Herrero A., Rieke M.J., Rieke G.H., Ruiz M., 1997, ApJ,
482, 747

\noindent Baars, J. W. M., Genzel, R., Pauliny-Toth, I. I. K., Witzel,
A., 1977, A\&A, 61, 99

\noindent Bower G.A., Wilson A.S., Mulchaey J.S., Miley G.K., Heckman T.M., Krolik J.H., 
1994, AJ, 107, 1686

\noindent Cole G.H.C., Pedlar A., Mundell C.G., Gallimore J.F,
Holloway A. J., 1998, MNRAS, 301, 782

\noindent Dopita M.A., Sutherland R.S. 1995,  ApJ, 455, 468.

\noindent Falcke H., Wilson A.S., Simpson C., 1998. ApJ, 502, 199.

\noindent Ferland G., Korista K.T., Verner D.A., Ferguson J.W., Kingdon
J.B., Verner E.M., 1998. PASP, 110, 761

\noindent Forbes D.A., Ward M.J. 1993,  ApJ, 416, 150. 

\noindent Hodapp K.W., Jensen J.B., Irwin E.M., Yamada H. Chung R. 
Fletcher K. Robertson L. Hora J.L., Simons D.A., Mays W.  2003,  PASP, 115, 1388. 

\noindent Horne K., 1986. PASP 98, 609

\noindent Mouri H., Kawara K., Taniguchi Y., 2000,  ApJ, 528, 186. 

\noindent Oliva E., Marconi A., Maiolino R., Testi L., Mannucci F., Ghinassi F., Licandro J., Origlia, L., Baffa, C., Checcucci, A.,  2001,  A\&A, 369L, 5. 

\noindent Pedlar A., Harrison B., Unger S.W., Graham D.A., Preuss E., Saikia D.J., Yates G.J. 1988,  LNP, 307, 310. 

\noindent Perryman M.A.C., 1997. The Hipparcos and Tycho catalogues.
Astrometric and photometric star catalogues derived from the ESA
Hipparcos Space Astrometry Mission, Noordwijk, Netherlands,
ESA Publications Division, 1997, ESA SP Series vol. 1200.

\noindent Press W.H., Teukolsky S.A., Vetterling W.T., Flannery B.P., 1992.
Numerical Recipes in C, Cambridge University Press, Cambridge.

\noindent Ramos Almeida C., P\'erez-Garcia A.M., Acosta-Pulido J.A.,
Rodr\'{\i}guez-Espinosa J.M., Barrena R., Manchado A., 2006. ApJ 645, 148

\noindent Riffel R., Rodr\'{\i}guez-Ardila A., Pastoriza M.G., 2006,
A\&A 457, 61

\noindent Rodr\'{\i}gues-Ardila A., Pastoriza M.G., Viegas S., Sigut
T.A.A., Pradhan A.K., 2004, A\&A, 425, 457

\noindent Rodr\'{\i}gues-Ardila A., Riffel R., Pastoriza M.G., 2005,
MNRAS, 364, 1041

\noindent Simpson, C. Forbes, D.A., Baker, A.C., Ward, M.J. 1996,  MNRAS, 283, 777. 

\noindent Su B.M., Muxlow T.W.B., Pedlar A., Holloway A.J., Steffen W., Kukula M.J., Mutel R.L. 1996,  MNRAS, 279, 1111. 

\noindent Sutherland R.S., Bicknell G.V., Dopita M.A. 2003,  ApJ, 591, 238. 

\noindent Ulvestad J.S., Wilson A.S. 1984,  ApJ, 278, 544. 

\noindent Whittle M., Wilson A.S. 2004,  AJ, 127, 606. 

\noindent Whittle M., Pedlar A., Meurs E.J.A., Unger S.W., Axon D.J., Ward M.J. 1988,  ApJ, 326, 125. 

\noindent Whittle M., Silverman J.D., Rosario D.J., Wilson
A.S.,  Nelson C.H., 2004, I.A.U. Symp. 222 ``The interplay among black holes, stars
and ISM in galactic nuclei'', Gramado, Brazil. March 2004,
Eds. T. Storchi-Bergmann, L. C. Ho and H. R. Schmitt. p299-302

\noindent Whittle M., Rosario D.J., Silverman J.D., Nelson C.H., Wilson A.S. 2005,  AJ, 129, 104.

\end{document}